\documentstyle[12pt]{article}
\newlength{\extraspace}
\newlength{\extraspaces}
\newcommand{\be}{\begin{equation}
\addtolength{\abovedisplayskip}{\extraspaces}
\addtolength{\belowdisplayskip}{\extraspaces}
\addtolength{\abovedisplayshortskip}{\extraspace}
\addtolength{\belowdisplayshortskip}{\extraspace}}
\newcommand{\ee}{\end{equation}}
\newcommand{\ba}{\begin{eqnarray}
\addtolength{\abovedisplayskip}{\extraspaces}
\addtolength{\belowdisplayskip}{\extraspaces}
\addtolength{\abovedisplayshortskip}{\extraspace}
\addtolength{\belowdisplayshortskip}{\extraspace}}
\newcommand{\ea}{\end{eqnarray}}
\newcommand{\nonu}{\nonumber \\[.5mm]}
\newcommand{\A}{&\!\!\!}
\begin{document}
\thispagestyle{empty}
\begin{flushright}
SIT-LP-06/02 \\
{\tt hep-th/0602219} \\
February, 2006
\end{flushright}
\vspace{7mm}
\begin{center}
{\large{\bf Spinor-vector supersymmetry algebra in three dimensions 
}} \\[20mm]
{\sc Kazunari Shima}
\footnote{
\tt e-mail: shima@sit.ac.jp} \ 
and \ 
{\sc Motomu Tsuda}
\footnote{
\tt e-mail: tsuda@sit.ac.jp} 
\\[5mm]
{\it Laboratory of Physics, 
Saitama Institute of Technology \\
Fukaya, Saitama 369-0293, Japan} \\[20mm]
\begin{abstract}
We focus on a spin-3/2 supersymmetry (SUSY) algebra 
of Baaklini in $D = 3$ 
and explicitly show a nonlinear realization of the SUSY algebra. 
The unitary representation of the spin-3/2 SUSY algebra 
is discussed and compared with the ordinary (spin-1/2) SUSY algebra. 
%
%
\end{abstract}
\end{center}

\newpage

\noindent
Supersymmetry (SUSY) algebra based on a spinor-vector generator, 
which is called the spin-3/2 SUSY, was first introduced by Baaklini \cite{Ba}, 
in which a nonlinear (NL) representation of the algebra has been realized 
by introducing a spin-3/2 Nambu-Goldstone (NG) fermion. 
This work was done by extending the NL realization 
of the ordinary (spin-1/2) SUSY algebra in terms of a spin-1/2 NG fermion 
known as the Volkov-Akulov (VA) model \cite{VA}. 
For the spin-1/2 SUSY algebra, the relation between the NL 
and the linear (L) SUSY \cite{WZ}, i.e., 
the algebraic equivalence of the NL SUSY VA action 
with various (renormalizable) spontaneously broken linear (L) supermultiplets 
has been investigated in \cite{IK}-\cite{STT}. 
According to this fact, corresponding L supermultiplets to a spin-3/2 NL SUSY 
action \cite{Ba} are also expected for the case of the spin-3/2 SUSY 
through a linearization. 

As a first step to investigate the L realization of the spin-3/2 algebra 
and the linearization of the spin-3/2 NL SUSY, 
we have recently studied the unitary representation of the spin-3/2 SUSY algebra 
of Baaklini \cite{ST3/2I} and a ($N = 1$) spin-3/2 L SUSY invariance 
of a free action in terms of spin-$(0, 1/2, 1, 3/2)$ fields \cite{ST3/2II}. 
\footnote{
We just mention the relation to the so-called no-go theorem \cite{CM,HLS} 
based upon the S-matrix arguments, 
i.e. the case for the S-matrix (the true vacuum) is well defined.  
As for the spin-3/2 L SUSY multiplet 
we have discussed the {\it global} L SUSY with spin-3/2 charges 
for the {\it free} action, which are free from the no-go theorem, so far. 
}
Those works are important preliminary not only to find out 
the (spontaneously broken) L SUSY supermultiplets which are equivalent 
to the NL realization of the spin-3/2 SUSY algebra, 
but also to obtain some informations for linearizing 
the {\it interacting global} NL SUSY theory with spin-3/2 (NG) fields 
in curved spacetime \cite{ST3/2SGM}. 
It may give new insight into an analogous mechanism 
with the super-Higgs one \cite{DZ} for high spin fields which appear 
in a composite unified theory based on SO(10) super-Poincar\'e (SP) group 
(the superon-graviton model) \cite{KS,ST}. 

In order to study further the spin-3/2 L supermultiplet structure, 
e.g., the closure on commutator algebras of L SUSY transformations, 
the structure of auxiliary fields, 
the mechanism of spontaneous SUSY breaking and the linearization of NL SUSY etc., 
it is useful to consider lower dimensional cases for simplicity of calculations. 
In this letter we focus on the spin-3/2 SUSY algebra of Baaklini 
in $D = 3$ and explicitly show the NL realization of the SUSY algebra. 
The unitary representation of the spin-3/2 SUSY algebra 
is further discussed and results are compared with those 
of the spin-1/2 SUSY algebra. 

Let us introduce the spin-3/2 SUSY (SP) algebra of Baaklini 
in $D = 3$ based on a spinor-vector generator $Q^a_\alpha$ as 
\footnote{
Minkowski spacetime indices are denoted 
by $a, b, \cdots = 0, 1, 2$ 
and two-component spinor indices are $\alpha, \beta = (1), (2)$. 
The Minkowski spacetime metric is 
${1 \over 2}\{ \gamma^a, \gamma^b \} = \eta^{ab} = {\rm diag}(+, -, -)$ 
and $\sigma^{ab} = {i \over 2}[\gamma^a, \gamma^b] 
= - \epsilon^{abc} \gamma_c$ $(\epsilon^{012} = -1 = \epsilon_{012})$, 
where we use the $\gamma$ matrices defined as $\gamma^0 = - \sigma^2$, 
$\gamma^1 = i \sigma^3$, $\gamma^2 = - i \sigma^1$ 
with $\sigma^I (I = 1, 2, 3)$ being Pauli matrices. 
We also use the charge conjugation matrix, $C = \sigma^2 = - \gamma^0$. 
}
\ba
\ [ P^a, P^b ] \A = \A 0, 
\label{3/2SUSY-PP}
\\
\ [ P^a, J^{bc} ] \A = \A i (\eta^{ab} P^c - \eta^{ac} P^b), 
\\
\ [ J^{ab}, J^{cd} ] 
\A = \A - i (\eta^{ac} J^{bd} - \eta^{bc} J^{ad} - \eta^{ad} J^{bc} + \eta^{bd} J^{ac}), 
\\
\ [ Q^a_\alpha, P^b ] \A = \A 0, 
\\
\ [ Q^a_\alpha, J^{bc} ] \A = \A {1 \over 2} (\sigma^{bc})_\alpha{}^\beta Q^a_\beta 
+ i (\eta^{ab} Q^c_\alpha - \eta^{ac} Q^b_\alpha), 
\nonu
\A = \A - {1 \over 2} \epsilon^{bcd} (\gamma_d)_\alpha{}^\beta Q^a_\beta 
+ i (\eta^{ab} Q^c_\alpha - \eta^{ac} Q^b_\alpha), 
\label{3/2SUSY-QJ}
\\
\{ Q^a_\alpha, Q^b_\beta \} 
\A = \A i \epsilon^{abc} (C)_{\alpha \beta} P_c, 
\label{3/2SUSY-QQ}
\ea
where $P^a$ and $J^{ab}$ are taranslational and Lorentz generators 
of the Poincar\'e group. 
It can be shown that Eqs.(\ref{3/2SUSY-PP})-(\ref{3/2SUSY-QQ}) 
satisfy all the Jacobi identities, in particular, the identities, 
\ba
\A \A 
[ Q^a_\alpha, [ J^{bc}, J^{de} ] ] 
+ [ J^{de}, [ Q^a_\alpha,  J^{bc} ] ] 
+ [ J^{bc}, [ J^{de}, Q^a_\alpha ] ] = 0, 
\nonu
\A \A 
\{ Q^a_\alpha, [ Q^b_\beta, J^{cd} ] \} 
+ [ J^{cd}, \{ Q^a_\alpha,  Q^b_\beta \} ] 
+ \{ Q^b_\beta, [ Q^a_\alpha, J^{cd} ] \} = 0, 
\ea
by means of the relation 
$\eta^{ab} \epsilon^{cde} = \eta^{ac} \epsilon^{bde} 
+ \eta^{ad} \epsilon^{cbe} + \eta^{ae} \epsilon^{cdb}$. 

As parallel discussions with \cite{Ba,VA}, 
a NL representation of the spin-3/2 SUSY algebra which reflects Eq.(\ref{3/2SUSY-QQ}) 
can be easily realized in terms of a Majorana spin-3/2 NG field $\psi^a$. 
Indeed, supertranslations of the $\psi^a$ 
and the Minkowski coordinate $x^a$ parametrized by a global Majorana 
spinor-vector parameter $\zeta^a$ are given by 
\ba
\A \A 
\delta \psi^a = \zeta^a, 
\nonu
\A \A 
\delta x^a = \kappa \epsilon^{abc} \bar\psi_b \zeta_c, 
\label{3/2st}
\ea
where $\kappa$ is a constant whose dimension is $({\rm mass})^{-3}$. 
Eq.(\ref{3/2st}) means NL SUSY transformation of $\psi^a$ 
at a fixed spacetime point as 
\be
\delta_Q \psi^a = \zeta^a 
- \kappa \epsilon^{bcd} \bar\psi_b \zeta_c \partial_d \psi^a, 
\label{3/2NLSUSY}
\ee
which gives the closed off-shell commutator algebra, 
\be
[ \delta_{Q1}, \delta_{Q2} ] = \delta_P(\Xi^a), 
\label{spin-3/2com}
\ee
where $\delta_P(\Xi^a)$ means a translation with a generator 
$\Xi^a = - 2 \kappa \epsilon^{abc} \bar\zeta_{1b} \zeta_{2c}$. 
Based upon a spin-3/2 NL SUSY invariant differential one-form defined as 
\ba
\omega^a \A \A 
= d x^a + \kappa \epsilon^{abc} \bar\psi_b d \psi_c 
\nonu
\A \A = (\delta^a_b 
+ \kappa \epsilon^{acd} \bar\psi_c \partial_b \psi_d) \ dx^b 
\nonu
\A \A = (\delta^a_b + t^a{}_b) \ dx^b 
\nonu
\A \A = w^a{}_b \ dx^b, 
\ea
an action, which is invariant under the spin-3/2 NL SUSY transformation (\ref{3/2NLSUSY}), 
is constructed as the volume form in $D = 3$: 
\ba
S = \A \A - {1 \over {2 \kappa}} 
\int \omega^0 \wedge \omega^1 \wedge \omega^2 
\nonu
= \A \A - {1 \over {2 \kappa}} 
\int d^3 x \ \vert w \vert 
\nonu
= \A \A 
- {1 \over {2 \kappa}} \int d^3 x 
\left[ 1 + t^a{}_a 
+ {1 \over 2!}(t^a{}_a t^b{}_b - t^a{}_b t^b{}_a) 
+ {1 \over 3!} \epsilon_{abc} \epsilon^{def} t^a{}_d t^b{}_e t^c{}_f 
\right], 
\label{3/2NLSUSYact}
\ea
where the second term, $-(1/2 \kappa) \ t^a{}_a 
= -(1/2) \epsilon^{abc} \bar\psi_a \partial_b \psi_c$, 
means the kinetic term for $\psi^a$. 

To know the L supermultiplet structure as corresponding one 
to the above spin-3/2 NL SUSY model, 
we study the unitary representation of the spin-3/2 SUSY algebra 
(\ref{3/2SUSY-PP})-(\ref{3/2SUSY-QQ}). 
By the use of similar methods to the case of spin-1/2 SUSY 
in $D = 3$ (for example, see \cite{HL}), 
operators which raise or lower the helicity of states 
for massive representations are constructed from the commutation relation 
(\ref{3/2SUSY-QJ}) as explains below: 
In fact, Eq.(\ref{3/2SUSY-QJ}) for the single helicity operator $J = J^{12}$ is 
\ba
\A \A [ Q^0_{(1)}, J ] = {i \over 2} Q^0_{(2)}, \ \ \ \ \ \ \ 
[ Q^0_{(2)}, J ] = - {i \over 2} Q^0_{(1)}, 
\nonu
\A \A [ Q^1_{(1)}, J ] = {i \over 2} Q^1_{(2)} - i Q^2_{(1)}, \ \ \ 
[ Q^1_{(2)}, J ] = - {i \over 2} Q^1_{(1)} - i Q^2_{(2)}, 
\nonu
\A \A [ Q^2_{(1)}, J ] = {i \over 2} Q^2_{(2)} + i Q^1_{(1)}, \ \ \ 
[ Q^2_{(2)}, J ] = - {i \over 2} Q^2_{(1)} + i Q^1_{(2)}. 
\label{QJcomp}
\ea
If we define the following operators 
as linear combinations of the real charges $Q^a_{(1)}$ and $Q^a_{(2)}$, 
\ba
\A \A 
R^{a \pm} = {1 \over 2} (Q^a_{(1)} \pm i Q^a_{(2)}), 
\ \ \ (R^{a \pm})^\dagger = {1 \over 2} (Q^a_{(1)} \mp i Q^a_{(2)}) = R^{a \mp}, 
\label{R+-}
\\
\A \A 
S^{\pm} = R^{1 \pm} + i R^{2 \pm} 
= {1 \over 2} \{ (Q^1_{(1)} \pm i Q^1_{(2)}) + i (Q^2_{(1)} \pm i Q^2_{(2)}) \}, 
\label{S+-}
\\
\A \A 
(S^{\pm})^\dagger = R^{1 \mp} - i R^{2 \mp} 
= {1 \over 2} \{ (Q^1_{(1)} \mp i Q^1_{(2)}) - i (Q^2_{(1)} \mp i Q^2_{(2)}) \}, 
\label{S+-dagger}
\ea
then commutators between the operators (\ref{R+-})-(\ref{S+-dagger}) 
and the $J$ become 
\ba
\A \A 
[ R^{0-}, J ] = - {1 \over 2} R^{0-}, 
\ \ \ [ R^{0+}, J ] = {1 \over 2} R^{0+}, 
\label{R+-J}
\\
\A \A 
[ S^+, J ] = - {1 \over 2} S^+, 
\ \ \ [ (S^+)^\dagger, J ] = {1 \over 2} (S^+)^\dagger, 
\label{S+daggerJ}
\\
\A \A 
[ S^-, J ] = - {3 \over 2} S^-, 
\ \ \ [ (S^-)^\dagger, J ] = {3 \over 2} (S^-)^\dagger. 
\label{S-daggerJ}
\ea
due to Eq.(\ref{QJcomp}). 
Eqs.(\ref{R+-J}) and (\ref{S+daggerJ}) 
mean that $R^{0\pm}$ and $(S^+, (S^+)^\dagger)$, 
raise or lower the helicity of states by 1/2, 
while Eq.(\ref{S-daggerJ}) shows that $(S^-, (S^-)^\dagger)$ 
raise or lower the helicity of states by 3/2. 

Furthermore, according to the anticommutation relation 
(\ref{3/2SUSY-QQ}), which is explicitly written in the component form as 
\ba
\A \A 
\{ Q^1_{(1)}, Q^2_{(2)} \} = - P_0, \ \ \ 
\{ Q^1_{(2)}, Q^2_{(1)} \} = P_0, 
\nonu
\A \A 
\{ Q^2_{(1)}, Q^0_{(2)} \} = - P_1, \ \ \ 
\{ Q^2_{(2)}, Q^0_{(1)} \} = P_1, 
\nonu
\A \A 
\{ Q^0_{(1)}, Q^1_{(2)} \} = - P_2, \ \ \ 
\{ Q^0_{(2)}, Q^1_{(1)} \} = P_2 
\ea
with all other anticommutators being zero, 
nonvanishing ones among the operators (\ref{R+-})-(\ref{S+-dagger}) are 
\ba
\A \A 
\{ S^+, (S^+)^\dagger \} = P_0, 
\label{S+S+dagger}
\\
\A \A 
\{ S^-, (S^-)^\dagger \} = - P_0, 
\\
\A \A 
\{ R^{0-}, S^+ \} = - {1 \over 2} (P_1 + i P_2), 
\label{R-S+}
\\
\A \A 
\{ R^{0+}, S^- \} = {1 \over 2} (P_1 + i P_2), 
\\
\A \A 
\{ R^{0-}, (S^-)^\dagger \} = {1 \over 2} (P_1 - i P_2), 
\\
\A \A 
\{ R^{0+}, (S^+)^\dagger \} = - {1 \over 2} (P_1 - i P_2). 
\label{R+S+dagger}
\ea
Note that $\{ R^{0-}, R^{0+} \} = 0$ for the $R^{0\pm}$ of Eq.(\ref{R+-J}). 
Obviously, by taking $P_a = (m, 0, 0)$ for the massive case $P^2 = m^2$ 
and by defining creation and annihilation operators, 
\ba
\A \A 
a_1 = {1 \over \sqrt{m}} S^+, \ \ \ 
a_1^\dagger = {1 \over \sqrt{m}} (S^+)^\dagger, 
\nonu
\A \A 
a_2 = {1 \over \sqrt{m}} S^-, \ \ \ 
a_2^\dagger = {1 \over \sqrt{m}} (S^-)^\dagger, 
\ea
Eqs.(\ref{S+S+dagger})-(\ref{R+S+dagger}) become 
\ba
\A \A 
\{ a_1, a^\dagger_1 \} = 1, \ \ \ 
\{ a_2, a^\dagger_2 \} = - 1, 
\nonu
\A \A 
\{ a_i, a_j \} = 0, \ \ \ \{ a^\dagger_i, a^\dagger_j \} = 0, 
\label{aa+}
\ea
where $i, j = 1, 2$. 
Eq.(\ref{aa+}) shows that only the $(a_1, a^\dagger_1)$ are 
the operators in the Fock space 
which shift the helicity of states by 1/2, 
while the $(a_2, a^\dagger_2)$ which shift by 3/2 
gives the negative norm (the non-unitary representation). 
Therefore, a (physical) massive irreducible representation 
for the spin-3/2 SUSY algebra in $D = 3$ 
induced only from $(a_1, a^\dagger_1)$ contains the stuffs with helicity 
$(\lambda, \lambda + 1/2)$, e.g., $(0, 1/2)$ as the simplest case. 
Namely, as for the physical modes of the massive multiplets in $D = 3$ 
has the same L supermultiplet structure as that of 
the $N = 1$ spin-1/2 SUSY in $D = 4$ \cite{WB}, 
although the multiplicity is another. 
Also they are compatible with the $N = 1$ spin-1/2 SUSY 
in $D = 3$ case \cite{HL} obtained in the different context. 

On the other hand, we are not able to construct consistent 
unitary representations for the massless case $P^2 = 0$, 
e.g., with a reference frame $P_a = (\epsilon, 0, \epsilon)$, 
since Eqs.(\ref{R-S+})-(\ref{R+S+dagger}) does not vanish. 
This is the similar situation to the spin-1/2 SUSY in $D = 3$, 
where the helicity in the massless case is no more good concept. 
In this case it may be necessary to use 
the method of constructing massless spin-1/2 $D = 3$ supermultiplets 
by extending (the representation space of) SUSY algebra \cite{dWTN}. 
We finally note that in $D = 2$ case of the spin-3/2 SUSY algebra 
the anticommutator $\{ Q^a_\alpha, Q^b_\beta \}$ of Baaklini's type vanishes. 

To summarize, we focus in this letter on the spin-3/2 SUSY algebra 
of Baaklini in $D = 3$ and show the NL realization 
of the SUSY algebra (\ref{3/2SUSY-PP})-(\ref{3/2SUSY-QQ}); 
namely, we explicitly construct the action (\ref{3/2NLSUSYact}) 
in terms of the spin-3/2 NG fermion $\psi^a$ 
which is invariant under the NL SUSY transformation (\ref{3/2NLSUSY}). 
We further discuss on the unitary representation 
of the spin-3/2 SUSY algebra in $D = 3$ 
towards the linearization of the spin-3/2 NL SUSY. 
For the massive case, the physical irreducible representation 
is induced only from the operators $(a_1, a^\dagger_1)$ 
which shift the helicity of states by 1/2 in Eq.(\ref{aa+}). 
On the other hand, no consistent unitary representation is constructed 
for the massless case by means of the nonvanishing anticommutation relations 
(\ref{R-S+})-(\ref{R+S+dagger}). 
This fact reflects the similar situation to the spin-1/2 SUSY in $D = 3$ 
with no good concept of helicity in the massless case, 
and we need further investigations to construct 
massless spin-3/2 $D = 3$ supermultiplets.

\newpage

%
\newcommand{\NP}[1]{{\it Nucl.\ Phys.\ }{\bf #1}}
\newcommand{\PL}[1]{{\it Phys.\ Lett.\ }{\bf #1}}
\newcommand{\CMP}[1]{{\it Commun.\ Math.\ Phys.\ }{\bf #1}}
\newcommand{\MPL}[1]{{\it Mod.\ Phys.\ Lett.\ }{\bf #1}}
\newcommand{\IJMP}[1]{{\it Int.\ J. Mod.\ Phys.\ }{\bf #1}}
\newcommand{\PR}[1]{{\it Phys.\ Rev.\ }{\bf #1}}
\newcommand{\PRL}[1]{{\it Phys.\ Rev.\ Lett.\ }{\bf #1}}
\newcommand{\PTP}[1]{{\it Prog.\ Theor.\ Phys.\ }{\bf #1}}
\newcommand{\PTPS}[1]{{\it Prog.\ Theor.\ Phys.\ Suppl.\ }{\bf #1}}
\newcommand{\AP}[1]{{\it Ann.\ Phys.\ }{\bf #1}}

\end{document}